# Structurally assisted melting of excitonic correlations in 1T-TiSe$_2$

*Max Burian[1,†], Michael Porer[1,†], Jose R. L. Mardegan[1,a], Vincent Esposito[1,b], Sergii Parchenko[1,c], Bulat Burganov[2], Namrata Gurung[3,4], Mahesh Ramakrishnan[1], Valerio Scagnoli[3,4], Hiroki Ueda[1], Sonia Francoual[5], Federica Fabrizi[6], Yoshikazu Tanaka[7], Tadashi Togashi[7,8], Yuya Kubota[7,8], Makina Yabashi[7,8], Kai Rossnagel[9,10], Steven L. Johnson[2,11] and Urs Staub[1,*]*

[1] Swiss Light Source, Paul Scherrer Institute, 5232 Villigen PSI, Switzerland.

[2] Institute for Quantum Electronics, ETH Zürich, Auguste-Piccard-Hof 1, 8093 Zürich, Switzerland.

[3] Laboratory for Mesoscopic Systems, Department of Materials, ETH Zurich, 8093 Zürich, Switzerland.

[4] Laboratory for Multiscale Materials Experiments, Paul Scherrer Institute, 5232 Villigen PSI, Switzerland.

[5] Deutsches Elektronen-Synchrotron DESY, Notkestraße 85, 22607 Hamburg, Germany.

[6] Diamond Light Source Ltd., Didcot, Oxfordshire OX11 0DE, United Kingdom.

[7] RIKEN SPring-8 Center, 1-1-1 Kouto, Sayo-cho, Sayo-gun, Hyogo 679-5148, Japan.

[8] Japan Synchrotron Radiation Research Institute (JASRI), 1-1-1 Kouto, Sayo-cho, Sayo-gun, Hyogo 679-5198, Japan.

[9] Institut für Experimentelle und Angewandte Physik, Christian-Albrechts-Universität zu Kiel, D-24098 Kiel, Germany.

[10] Ruprecht-Haensel-Labor, Deutsches Elektronen-Synchrotron DESY, D-22607 Hamburg, Germany.

[11] SwissFEL, Paul Scherrer Institute, 5232 Villigen PSI, Switzerland.

[a] current address: Deutsches Elektronen-Synchrotron DESY, Notkestraße 85, 22607 Hamburg, Germany.

[b] current address: Stanford Institute for Materials and Energy Sciences, SLAC National Accelerator Laboratory, 2575 Sand Hill Road, Menlo Park, CA 94025.

[c] current address: Laboratory for Mesoscopic Systems, Department of Materials, ETH Zurich, 8093 Zürich, Switzerland.

[†] These authors contributed equally.

[*] To whom correspondence should be addressed. Email: urs.staub@psi.ch

## Abstract

The simultaneous condensation of electronic and structural degrees of freedom gives rise to new states of matter, including superconductivity and charge-density-wave formation. When exciting such a condensed system, it is commonly assumed that the ultrafast laser pulse disturbs primarily the electronic order, which in turn destabilizes the atomic structure. Contrary to this conception, we show here that structural destabilization of few atoms causes melting of the macroscopic ordered charge-density wave in 1T-TiSe$_2$. Using ultrafast pump-probe non-resonant and resonant X-ray diffraction, we observe full suppression of the Se 4p orbital order and the atomic structure at excitation energies more than one order of magnitude below the suggested excitonic binding energy. Complete melting of the charge-density wave occurs 4-5 times faster than expected from a purely electronic charge-screening process, strongly suggesting a structurally assisted breakup of excitonic correlations. Our experimental data clarifies several questions on the intricate coupling between structural and electronic order in stabilizing the charge-density-wave in 1T-TiSe$_2$. The results further show that electron-phonon-coupling can lead to different, energy dependent phase-transition pathways in condensed matter systems, opening new possibilities in the conception of non-equilibrium phenomena at the ultrafast scale.

# Introduction

The interplay of electronic and structural order creates a rich playground for the manipulation of properties of condensed matter systems. Indeed, many electronic material properties, such as the macroscopic spin state [1–3] and conductivity [4–6], have a strong, often non-linear dependence on the structural order, represented by local displacements of the ions. This dependence is often ascribed to strong electron-phonon coupling [7,8], where tiny changes in atomic positions significantly alter the electronic landscape and vice versa [9–11]. Disentangling the intrinsic electronic from the structurally induced contributions is hence a complex task. It requires specificity to both orders at a time and energyscale where charge- and phonon-driven processes occur: the (ultrafast) femtosecond regime [1–3,5,7–9,12].

A prominent yet controversial system in which the interplay of electronic and structural order is of particular interest is the transition-metal dichalcogenide, titanium diselenide (1T-TiSe$_2$). Apart from exhibiting doping [13] and pressure [14] induced superconductivity, it shows a commensurate charge-density wave (CDW) at $T_c \approx 200K$ [15–17], proposed as a result of Bose-Einstein condensation of excitons [18] and potentially chiral [19], gyrotropic electronic order [20]. In its room-temperature (RT) phase, 1T-TiSe$_2$ has its electron (Ti) and hole pockets (Se) at the L and Γ points of the Brillouin zone, respectively [21–23]. Valence and conduction bands at these Fermi-pockets present a small (negative) band gap [22,24], causing charge-carrier accumulation [25] and hence an overall semimetal-like material character [22,26,27] (see Figs. 1a and 1b). Upon cooling below $T_c$, the initial lattice ($P\bar{3}m1$ space-group) undergoes a periodic lattice deformation (PLD) [28,29] leading to a doubling of the unit cell (see Fig. 1c) into the (2 x 2 x 2) superlattice CDW phase ($P\bar{3}c1$ space-group) [16,28,30]. This structural doubling folds the Fermi-pockets on top of each other, leading to a partial opening of the (p-d) gap at L/Γ* (see Figs. 1a and 1b) [31–33] and local distortion of the hybridized orbitals [34]. In short: the phase transition in 1T-TiSe$_2$ hence occurs on (i) the electronic, orbital order connected to the CDW and (ii) the structural order relating to the PLD.

While the general picture of the CDW phase transition may be understood, its mechanism remains highly controversial. One proposed mechanism involves thermally allowed condensation of free carriers (holes and electrons) to bound excitons that destabilize the room-temperature crystal structure, in turn causing the PLD [23,30,35,36]. Indeed, charge screening of excitons by photoinjected carriers has been shown to melt the CDW phase and the associated band-folding [31,37], revealing the necessity of excitonic correlations in stabilizing the CDW. Another mechanism is suggested by recent calculations, indicating that the PLD is required to both energetically stabilize excitonic correlations in the CDW state [38] and to reproduce the electronic band structure [34]. Moreover, the observed mixing of phonon and plasmon bands at $T_c$ is a strong evidence for electron-phonon-coupling [9,18], pointing towards a prominent role of structural contributions [39,40]. Undoubtedly, the PLD is of unambiguous importance for CDW stabilization, [41] although the extent of its influence is unclear. In particular there is an outstanding question of how important the PLD is for exciton condensation.

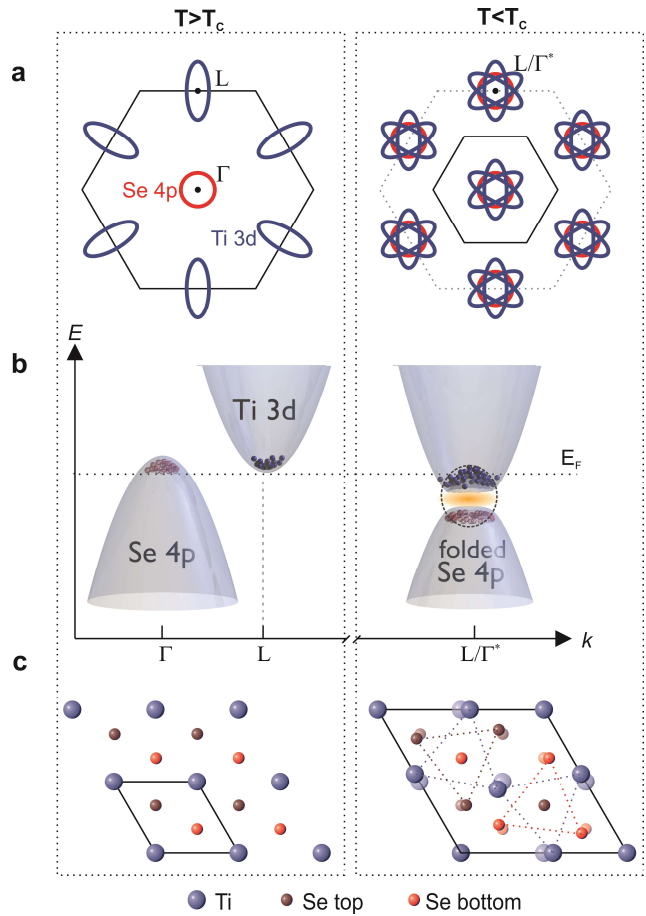

**Fig 1. CDW phase and associated PLD upon cooling of 1T-TiSe$_2$. a,** Simplified Fermi surface of a slice through Fermi pockets of holes (red – Se 4p) and electrons (blue – Ti 3d) in the first Brillouin zone at ±A along the $k_c$-axis, shown in the room-temperature (left) and CDW (right) phase. The band-folding of Ti 3d and Se 4p states accompanying the CDW transition is indicated, leading to a re-referencing of the Brillouin zone (e.g. the RT L point becomes CDW L/Γ$^*$ point). **b,** The 3D representation of the bands illustrates the p-d splitting between the folded Se 4p and Ti 3d bands at L/Γ$^*$, leading to excitonic correlations between electron-hole pairs (black dotted circle). **c,** PLD associated with the CDW phase, causing a doubling of the unit cell along all three dimensions.

Here, resonant X-ray diffraction offers a unique capability: by tuning the X-ray energy to a core-to-valence electron transition, the polarization dependent scattering cross-section is highly sensitive to occupation, geometry and overall-spin of the corresponding valence orbital [42–46]. This becomes especially useful when paired with the ultrashort pulses produced by X-ray free-electron lasers, where an optical laser pump modified the electronic landscape and a subsequent X-ray pulse probes the change in either orbital [47–49] or structural order [47,50–52], depending on the energy (resonant/non-resonant) or the type of reflection (e.g., charge-symmetry-forbidden but orbital-allowed). In this work, we use transient resonant/non-resonant X-ray diffraction after near-infrared photoexcitation to disentangle the contributions of structural and orbital order on the melting of the CDW in 1T-TiSe$_2$. We find an energy regime in which the PLD is destabilized while excitonic correlations persist: melting of the CDW is observed only when surpassing a critical activation energy barrier. The suppression of both structural and orbital order, occur significantly faster and at drastically lower threshold energies than expected for a pure charge screening induced breakup of excitonic correlations. These results suggest a structurally enhanced melting of the CDW phase, where destabilization of the PLD structurally assists the breakup of excitonic correlations.

**Experimental Methods**

The resonant/non-resonant X-ray diffraction experiment was carried out at the EH2 endstation of the BL3 beamline at the SACLA X-ray free electron laser [53]. An illustration of the experimental setup is shown in Fig. 2a. The sample, an ultra-thin TiSe$_2$ single crystal (c-axis out-of-plane, 61.9 ± 0.5 nm thickness – see Supplementary Fig. S1 and Supplementary Chapter SC1 for static sample characterization) in the CDW phase was mounted on a goniometer with a N$_2$-cryostream to stabilize the temperature at approximately 100 K. A slightly focused (300 x 300 μm) 35 fs optical pulse with wavelength 800 nm excited the sample with a repetition rate of 15 Hz. The horizontal polarized X-ray beam operating at 30 Hz was focused to a spot size of 10 x 10 μm$^2$. During the experiment, the angle between the X-ray- and laser-beams was kept fixed. To ensure identical excitation conditions, the diffraction geometries for structural and orbital reflection were kept as similar as possible. For the structural reflection, X-rays entered at a grazing angle of 1° with a sample orientation relative to the horizontal X-ray polarization of $\chi$ = 17°, corresponding to a laser incidence angle of 11°. For the orbital reflection, X-rays entered at a grazing angle of 2° with a sample orientation perpendicular to the horizontal X-ray polarization ($\chi$ = 90°), corresponding to a laser incidence angle of 12°. As the sample orientation was different for both reflections, a half-wave-plate was used to adjust the polarization of the optical laser to ensure fully p-polarized light at the different diffraction geometries. Notably, the effective X-ray penetration depth is significantly larger than the sample thickness of 62 nm, ensuring that the entire sample cross-section is probed. As only every second X-ray pulse follows laser excitation, we used shot-to-shot background correction to reduce effects of fluctuations in X-ray pulse intensity. The transients shown in this work come from 2D integration of the scattering peak on the detector, corrected for the fluorescence background. For orbital and structural reflection we did not observe a shift of the peak maximum but only a decrease in the scattering intensity. The X-ray beam energy was tuned in the vicinity of the Se K-edge around 12.65 keV (see Supplementary Fig. S2 for energy scan of the (0 0 5/2) reflection intensity taken at SACLA). The inevitable temporal jitter between X-ray and optical laser pulse was measured shot-by-shot using a transmission-grating based timing tool, which has an accuracy <10 fs (FWHM) [54]. The temporal fingerprint of each shot was then used to re-bin all data into 30 fs time-segments [55]. In summary, the effective time-resolution of the experiment $\Delta t_{Eff}$ has to take into account (i) the optical laser pulse width $\tau_L$ = 35 fs, (ii) the X-ray laser pulse width $\tau_{Xray}$

= 15 fs, (iii) the time resolution of the timing tool $\Delta t_{TT}$ = 10 fs and (iv) the size of data-bins $\Delta t_{bin}$ = 30 fs, leading to $\Delta t_{Eff} = \sqrt{\tau_L^2 + \tau_{xray}^2 + \Delta t_{TT}^2 + \Delta t_{bin}^2} \approx$ 50 fs.

In the experiment, the optical (800 nm) laser pulse drives an electronic inter-band transition in the TiSe$_2$ sample that is in the CDW phase. The excitation injects valence band (Se-VB) electrons from the folded Se 4p orbital as hot-carriers into the Ti 3d conduction band (Ti-CB). This hot-carrier injection occurs during the pump-pulse duration [31] (here < 35 fs FWHM). Shift in plasmon resonance frequency [56], bulk-lattice vibrations [50,56,57] or re-normalization of the Ti-CB [37] are observed at slower (50-500 fs) and experimentally observable timescales but depend on the incident laser pump fluence. Indeed, and as we have shown in previous work [56], pump-fluences above 0.3 mJ/cm$^2$ lead to a complete melting of both excitonic and structural order, driven by Coulomb screening-induced exciton breakup. For lower fluences (<0.3 mJ/cm$^2$), the excited electron plasma is strongly out-of-equilibrium while a fingerprint of PLD-induced back-folded acoustic phonon branches appears unaffected [56]. This fluence regime in which electronic and structural order deviate sets the ideal conditions to disentangle the relationship between CDW and PLD.

## Results

We capture the electronic response following excitation by selecting a (0 k/2 l/2) type reflection at the Se K-edge resonance (12.652 keV; see Supplementary Fig. S2). These reflections are space-group-forbidden in the RT and CDW phase for Thomson scattering. They are hence sensitive to the Se 4p orbital order, as shown by (i) the resonance behavior at the Se K-edge energy (see Supplementary Fig. S2), (ii) the presence of scattering intensity only in the rotated polarization channel (see Supplementary Fig. S3) and (iii) the absence of scattering above T$_c$ [58]. More precisely, the phase of the scattering amplitudes from all spherical/symmetric Se 4p orbital contributions cancels out, while all non-congruent, aspheric contributions (here at the $P_{\bar{3}c1}$ Se1 site at Wyckoff position *12g* [59]) remain present, giving rise to the diffraction signal (more details can be found in ref. [58]). As this Se 4p orbital asphericity is caused by site-specific hole condensation [34], this type of resonant reflections directly measures the CDW-related condensation of excitonic correlations.

Figure 2b shows the transient X-ray diffraction (tr-XRD) intensities of the (0 1/2 7/2) orbital reflection at the Se K-edge. Following the impulsive laser excitation, the diffraction intensity decreases rapidly within < 200 fs, reaching almost a complete suppression for an absorbed fluence of 45 μJ/cm$^2$ (see last trace in Fig. 2b) [60]. For fluences below 16 μJ/cm$^2$, a recovery on the timescale of approximately > 1 ps follows the rapid intensity decrease. For stronger excitations the intensity remains suppressed over the experimental time window – a phenomenon characteristic for phase transitions. [61,62] To quantify these experimental observations, we perform a global fit of the time-dependent order-parameter model [48,61], where the order parameter η relates to aspheric contributions of the Se 4p orbital geometry (for details see Supplementary Chapter SC2: Fitting of orbital order transients). This orbital asphericity parameter η reduces to zero if the absorbed local energy density per volume $n(t)$ exceeds a critical threshold $n_{c,Orb}$, such that the orbital symmetry for $n(t)/n_{c,Orb} > 1$ is the same as found in the RT phase where no excitonic correlations exist. For energy density depth profiles corresponding to selected traces see right panel of Fig. 2b.

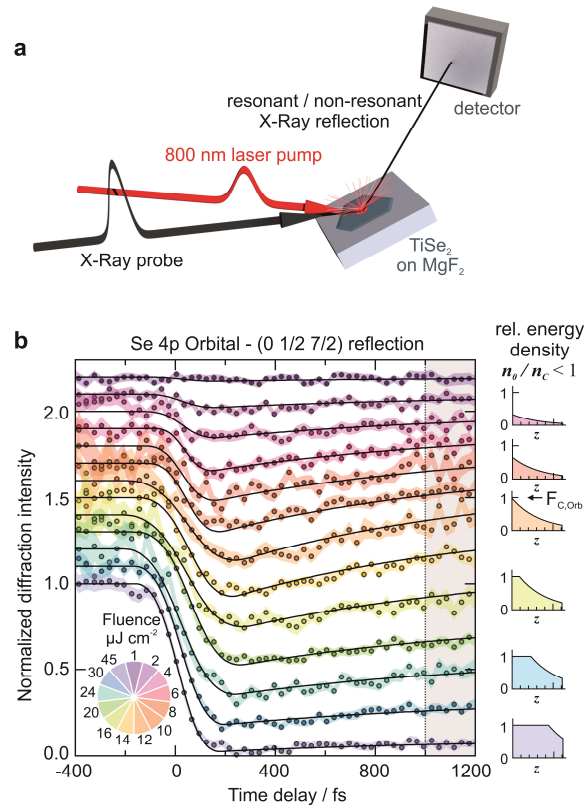

**Fig 2. Experimental overview and XRD transients of the orbital reflection**. **a**, Schematic representation of the experimental setup. **b**, Normalized diffraction intensities of the (0 1/2 7/2) reflection at the Se K-edge resonance (see colored markers for increasing fluence – curves have a constant offset for fluences below 45 µJ/cm² for better visibility). Black traces correspond to the orbital order parameter η from the global model-fit, based on the absorbed energy profiles at t=0 shown on the right. The gray area (1 < t < 1.2 ps) denotes the time-region for the analysis in Fig. 4a.

The experimental data and the model are in excellent agreement (see Fig. 2b and Supplementary Fig. S4), yielding three significant model parameters. We find an effective optical absorption length of 34.7 ± 1.3 nm compared to the literature value of 17 nm [63] (see Supplementary Fig. S5 for a model refined using the literature value). The discrepancy here is likely due to rapid ballistic transport of hot-carriers [37] or possibly non-linear absorption of the 35 fs (FWHM) pump pulse. The model further yields an exponential recovery constant of $\tau_{rel,Orb} = 1.08 \pm 0.03$ ps, which is in good agreement with electronic thermalization processes [57]. Third, we find complete suppression of the orbital order in the top most layer for fluences above 12.3 ± 0.5 µJ/cm² (**F$_{c,Orb}$**). Assuming an exponential absorption profile of the laser excitation, this implies that fluences > 33 µJ/cm² are required to suppress the orbital order of more than 50% of the sample volume. This compares well with the previously (not depth-resolved) determined threshold of approximately 40 µJ/cm² for switching to normal state conductivity caused by breakup of excitonic correlations [56,57]. This indicates that the observed melting of the Se 4p orbital order is linked to the reduction of the excitonic correlated nature of bound holes as breakup of excitonic correlations suppresses hole-induced orbital distortion and hence reduces the aspheric scattering contribution [64]. Most strikingly, the critical fluence **F$_{c,Orb}$** corresponds to a local energy threshold to melt the excitonic order of 1.35 ± 0.05 meV per Ti atom, which is significantly below theoretical estimates of the exciton binding energy in 1T-TiSe$_2$ (17 meV) [24,31] even when accounting for the thermal offset during the experiment (100 K, corresponding to 8.6 meV).

To understand whether the drastic reduction of the energy-barrier is linked to the structural order, we also acquired non-resonant (12.620 keV) transient tr-XRD traces for a CDW space-group allowed [(h/2 k/2 l/2) type] reflection (see Supplementary Fig. S2). It is sensitive to contributions of both A1g* and Eg* modes that soften at the L point during the RT- to CDW-phase transition [65–67], such that the diffraction intensity is zero if the PLD is suppressed.

Figure 3a shows the tr-XRD intensities of the (5/2 1/2 3/2) structural peak after excitation. Similar to the orbital reflection shown in Fig. 2b, the scattering signal decreases rapidly within <200 fs, reaching near complete suppression for fluences at 40 µJ/cm² (see bottom trace in Fig. 3a). Also here, the scattering intensity partially recovers for fluences below 10-16 µJ/cm², while a slight oscillatory component appears in the 0-400 fs time-window for fluences >25 µJ/cm². With a further increase in excitation energy above 100 µJ/cm² (see Fig. 3b) the oscillatory contribution both increases and becomes more coherent, while shifting in frequency from 2.5 ± 0.2 to 3.3 ± 0.2 THz for 107 and 227 µJ/cm², respectively. While these modes appear to be in good agreement with the Eg* (2.2 THz) and A1g* (3.4 THz) CDW phonon modes [50,57,65], the mode hardening with increasing fluence appears counterintuitive, especially since the equilibrium CDW- to RT-phase transition is accompanied by a significant mode-softening [66] and this reflection mainly probes the in-plane PLD component [58].

To describe the structural dynamics in the low (<45 µJ/cm²) and high (>100 µJ/cm²) fluence regime, we formulate a theoretical model based on a double-well-potential (DWP) [47,68] that describes the A1g* potential energy landscape, as depicted by the black trace in the right panel of Fig. 3. Prior to excitation, the system resides at the minimum positions corresponding to the equilibrium PLD in the CDW phase. Hot-carrier injection through laser excitation causes an instantaneous, transient change in the ionic potential energy landscape [9], resulting in coherent atomic oscillations around the the new, shifted potential minimum (see Supplementary Chapter SC3: Modelling of structural order transients for a detailed description). Following the initial excitation, the potential energy landscape relaxes slowly back to its original shape with an exponential time constant $\tau_{rel,Str}$.

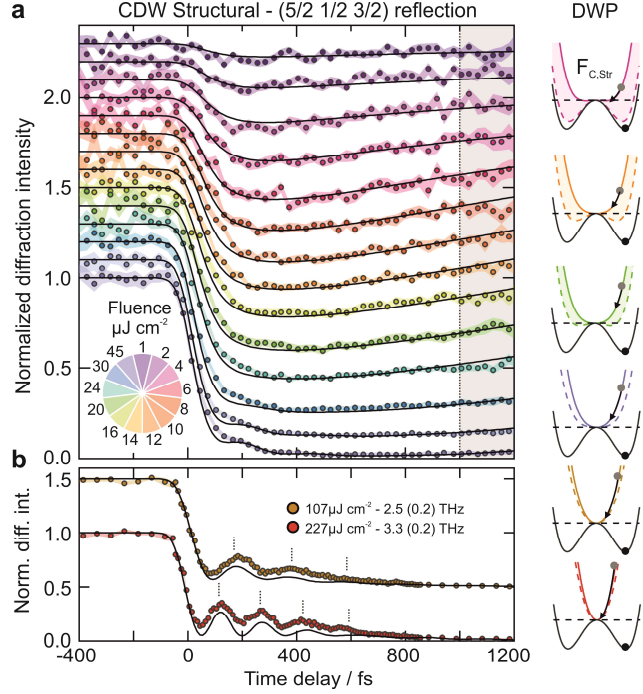

**Fig 3. XRD transients of the structural reflection**. **a**, Normalized diffraction intensities of the (5/2 1/2 3/2) reflection (see colored markers) with increasing fluence. The gray area denotes the time-region for the analysis in Fig. 4a. **b**, Diffraction intensities after high-fluence excitation, showing a strong oscillatory component (dashed markers note half-period). For **a** and **b**, black traces correspond to transients calculated from the double-well-potential model. On the right, energy potentials for selected traces at t=0 are shown (black: ground-state, solid color: excited potential in top layer, dashed-color: excited potential at backside of sample).

Again, the model agrees well with the tr-XRD data obtained experimentally (see Fig. 3). To be consistent with the fit to the orbital reflection transients, we fix the effective optical absorption length to 34 nm [69]. The exponential recovery parameter $\tau_{rel,Str}$ is 1.6 ps, which is slightly longer than the 1.08 ps obtained from the analysis of the orbital reflection transients and compares well to the relaxation of coherent structural dynamics in similar systems [47]. Further, we find suppression of the ground-state potential barrier for local energy densities > 0.52 meV per Ti-atom, which is reached in the top-most layer for fluences > 4.8 µJ/cm² ($F_{c,Str}$) (see pink schematic in the right panel of Fig. 3) [i.e., when the transient equilibrium coordinate of the ($P\bar{3}c1$) CDW amplitude mode becomes zero]. Further increase in fluence leads to a larger sample fraction with suppressed energy barrier and a further steepening of the potential-barrier wall for strongly excited regions near the surface. The lattice, following the landscape given by the single-well potential, now traverses through the high-temperature structure and overshoots, yielding an oscillating, CDW-like distorted lattice with an inverting amplitude at each zero crossing. We observe the oscillation at twice its frequency as the diffraction intensity scales with the *squared* amplitude. This is a coherent oscillation of a CDW-related $L_1$ soft-phonon [65] of the

RT ($P\bar{3}m1$) structure (Fig 3b). As the potential walls become steeper with increasing fluence, this $L_1$-mode-related oscillation effectively hardens – a transient analogue to the thermal behavior when temperature increases above $T_c$. [65,66]

## Discussion

The comparison of results obtained from the on- and off-resonance transient data suggests two distinct pathways for structural and orbital order, as evidenced by the different critical thresholds of 0.52 and 1.35 meV/Ti-atom, respectively. This difference is found not only in the quantitative model parameters, but also becomes directly apparent in the experimental data, particularly in the 1-1.2 ps time window (see gray areas in Figs. 2b and 3). In this time-regime (i) oscillatory contributions in the structural response have vanished and (ii) the orbital response is mainly sensitive to the sample fraction with energy densities above $F_{c,Orb}$, as contributions from lower densities have decayed by 1/e. As apparent in Fig. 4a, the scattering intensity of the structural reflection (red dots) decreases linearly with the pump fluence up to approximately 20 µJ/cm² (see transition region in inset of Fig. 4a). In contrast, fluence-dependence of the intensity of the orbital reflection (black dots) changes from a linear (see Fig. 4a blue dashed line for projection) into a super-linear correlation, with the pivot-point coinciding with the orbital threshold $F_{c,Orb}$. Remarkably, at this fluence the structural signal has already decreased by approx. 50%, in agreement with its significantly lower structural threshold $F_{c,Str}$. It is hence evident that at energy densities below 1.35 meV/Ti-atom, the PLD is significantly destabilized while excitonic correlations remain mostly intact.

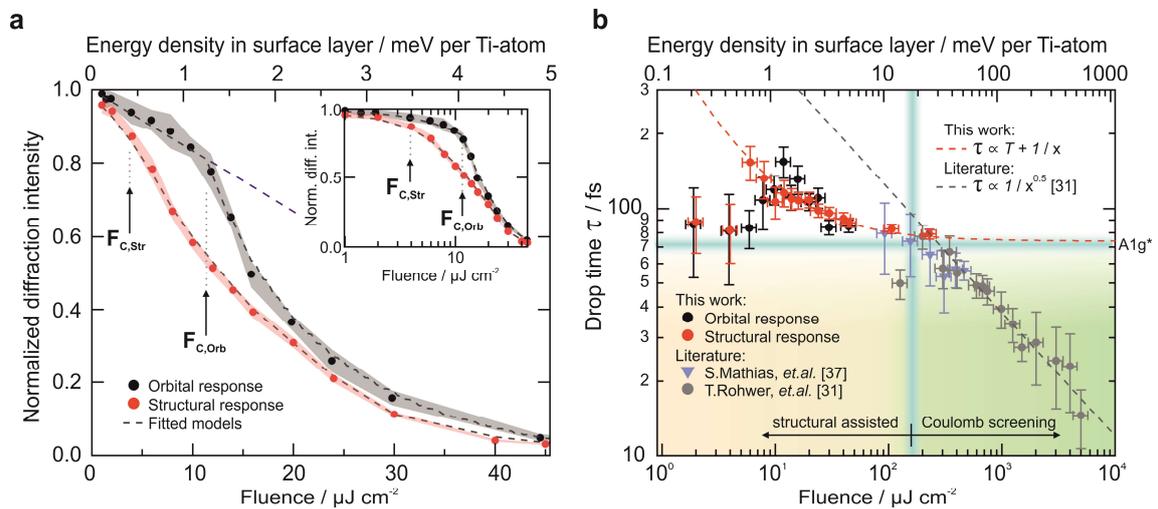

**Fig 4. Fluence dependence of orbital and structural order**. **a**, Averaged tr-XRD intensity of orbital (black) and structural (red) order in the 1.0 - 1.2 ps time window (see gray region in Figs. 2b and 3a). The blue dashed line indicates deviation of the orbital response from the linear fluence dependency, marked by orbital threshold energy $F_{c,Orb}$. The inset shows a logarithmic representation to highlight the transition in structural response at $F_{c,Str}$. **b**, Comparison of the dynamics of orbital and structural response with literature [31,37]. In the high-fluence regime (>200 µJ/cm²), hot-carrier screening of excitonic correlations is sufficiently fast to dominate the overall melting

dynamics. In the low-fluence regime (< 200 µJ/cm$^2$), structurally assisted destabilization following the soft-mode outpaces charge-screening and hence becomes the governing mechanism in melting of the CDW phase.

These results clearly point toward a structurally assisted melting of excitonic correlations. Hot-carrier or hole injection distorts the local energetic landscape [9], causing a structural rearrangement along the A1g* toward the RT phase. Note that this structural rearrangement also occurs for fluences below the orbital threshold $F_{c,Orb}$, in a state where excitonic correlations are not sufficiently affected by the pump. As the exciton condensate is not stable in the absence of a PLD, the structural movement toward the RT phase lowers the activation barrier for exciton breakup towards the observed threshold through altering the electronic band structure [34]. Surpassing this activation barrier, the melting of excitonic correlations initiates a self-amplifying cascade of band-edge renormalization and self-injection through hot-carrier relaxation [37], as evidenced by the super-linear fluence dependence of the orbital order (see Fig. 4a).

Indeed, this scenario, in which the absence of the PLD destabilizes excitonic correlations, is plausible only if the transient structural motion is sufficiently fast compared to the competing phenomenon of Coulomb-screening [9,31,37]. A comparison of the signal drop time τ of our data with the time constants for the transients of the folded Se 4p occupation determined by ARPES [31,37] is shown in Fig. 4b (see Supplementary Fig. S8). Obviously, the orbital and structural responses in the low-fluence regime are 4-5 times faster than the CDW melting processes, which are solely due to Coulomb screening induced exciton breakup (see that gray dotted line in Fig. 4b for Fröhlich-law prediction lies above our experimental data). All transients we observer are slower than 75 fs, which corresponds to a quarter oscillation of the A1g* soft-mode (that is argued to be the bottleneck to switch the lattice from CDW to RT phase [57]). These two findings strongly suggest that structural destabilization of the PLD is the dominant, rate-limiting process in the low-fluence regime. Interestingly, the intersection point of these two regimes (blue break lines in Fig. 4b) that limit the dynamics, charge screening versus soft-mode, locates precisely at the exciton binding energy of 17 meV [24]. This strongly supports the notion that two competing phenomena are responsible for CDW melting. For fluences above the exciton binding energy, charge-screening is sufficiently fast to outpace the phononic response [31,37], such that the corresponding THz fingerprint remains present despite the absence of excitonic correlations [56]. Indeed, for the highest fluences tested, the coherent oscillations of the structural order with the CDW-associated soft-mode implies that the periodicity associated with the PLD remains existing in an oscillatory state before the mode is damped out and the corresponding scattering intensity finally approaches zero. This may explain the observation of persistent back-folded acoustic phonon branches in the high excitation regime [56] despite the presence of a structural single-well potential. For fluences below the exciton binding energy, the structural response destabilizes excitonic correlations [39,40] and lowers the effective breakup barrier, leading to structurally assisted melting of the CDW phase.

# Conclusion

Our work provides experimental evidence that the PLD is not only a side effect but is instead a crucial ingredient in stabilizing the CDW phase in 1T-TiSe$_2$, as relaxation of the PLD lowers the effective dissociation threshold of excitonic correlations by more than one order of magnitude. It is further evident that the kinetics of competing processes seem to determine whether Coulomb screening from injected hot carriers or structural destabilization of the PLD triggers the overall melting of the CDW phase. Yet regardless of this initial trigger, our results show that both electronic and structural order melt quasi-simultaneously, confirming the presence of strong electron-phonon coupling in the 1T-TiSe$_2$ system [9].

For a detailed, microscopic understanding of the laser-driven phase transition, it may be necessary to consider the length scales at which different (dis)ordering processes originate: while laser excitation might not be sufficiently strong to drive a *macroscopic* electronic transition (i.e., Coloumb screening), it may still distort the *local* electronic potential that stabilizes the PLD, thereby initiating an expanding cascade of electron-phonon-coupling mediated transitions as a macroscopic effect. The crucial aspect of this mechanistic comprehension lies in the highly non-linear nature of the coupling interactions, which in our case lowers the effective binding energy of electronic correlations by more than one order of magnitude. We anticipate similar, potentially even more drastic, effects within the wide range of compounds that present strong coupling between different orders, such as, e.g., cuprates [4,10,70], other dichalcogenides [6,71,72], (complex) oxides [1,3,73] as well as superlattice [74,75] and spin-ice [76,77] systems.

# Acknowledgements

Static X-ray characterization of the sample was performed at the P09 beamline (PETRA III) and X04SA beamline (SLS) during in-house access as well as the I16 beamline (Diamond Light Source) under proposal numbers MT15742. The time-resolved x-ray experiments were performed under the approval of the Japan Synchrotron Radiation Research Institute (JASRI Proposal No.2017B8039). We thank C. Monney and T. Jaouen for enlightening discussions. The research leading to these results has received funding from the Swiss National Science Foundation and its National Centers of Competence in Research, NCCR MUST and NCCR MARVEL. This research was supported by the Swiss National Science Foundation (Grant No. 200021_169698).

# References

[1] K. W. Kim, A. Pashkin, H. Schäfer, M. Beyer, M. Porer, T. Wolf, C. Bernhard, J. Demsar, R. Huber, and A. Leitenstorfer, *Ultrafast Transient Generation Of Spin-Density-Wave Order In The Normal State Of $BaFe_2As_2$ Driven By Coherent Lattice Vibrations*, Nat. Mater. **11**, 497 (2012).

[2] C. Dornes, Y. Acremann, M. Savoini, M. Kubli, M. J. Neugebauer, E. Abreu, L. Huber, G. Lantz, C. A. F. Vaz, H. Lemke, et al., *The Ultrafast Einstein–De Haas Effect*, Nature **565**, 209 (2019).

[3] R. V. Mikhaylovskiy, E. Hendry, A. Secchi, J. H. Mentink, M. Eckstein, A. Wu, R. V. Pisarev, V. V. Kruglyak, M. I. Katsnelson, T. Rasing, et al., *Ultrafast Optical Modification Of Exchange Interactions In Iron Oxides*, Nat. Commun. **6**, 8190 (2015).

[4] D. Fausti, R. I. Tobey, N. Dean, S. Kaiser, A. Dienst, M. C. Hoffmann, S. Pyon, T. Takayama, H. Takagi, and A. Cavalleri, *Light-Induced Superconductivity In A Stripe-Ordered Cuprate*, Science **331**, 189 (2011).

[5] M. Rini, R. Tobey, N. Dean, J. Itatani, Y. Tomioka, Y. Tokura, R. W. Schoenlein, and A. Cavalleri, *Control Of The Electronic Phase Of A Manganite By Mode-Selective Vibrational Excitation*, Nature **449**, 72 (2007).

[6] B. Sipos, A. F. Kusmartseva, A. Akrap, H. Berger, L. Forró, and E. Tutiš, *From Mott State To Superconductivity In $1T$-$TaS_2$*, Nat. Mater. **7**, 960 (2008).

[7] H. Seiler, S. Palato, C. Sonnichsen, H. Baker, E. Socie, D. P. Strandell, and P. Kambhampati, *Two-Dimensional Electronic Spectroscopy Reveals Liquid-Like Lineshape Dynamics In $CsPbI_2$ Perovskite Nanocrystals*, Nat. Commun. **10**, 4962 (2019).

[8] K. Miyata, D. Meggiolaro, M. Tuan Trinh, P. P. Joshi, E. Mosconi, S. C. Jones, F. De Angelis, and X. Y. Zhu, *Large Polarons In Lead Halide Perovskites*, Sci. Adv. **3**, e1701217 (2017).

[9] C. Lian, S. J. Zhang, S. Q. Hu, M. X. Guan, and S. Meng, *Ultrafast Charge Ordering By Self-Amplified Exciton–Phonon Dynamics In $TiSe_2$*, Nat. Commun. **11**, 43 (2020).

[10] A. Lanzara, P. V. Bogdanov, X. J. Zhou, S. A. Kellar, D. L. Feng, E. D. Lu, T. Yoshida, H. Eisaki, A. Fujimori, K. Kishio, et al., *Evidence For Ubiquitous Strong Electron-Phonon Coupling In High-Temperature Superconductors*, Nature **412**, 510 (2001).

[11] E. Pastor, J. S. Park, L. Steier, S. Kim, M. Grätzel, J. R. Durrant, A. Walsh, and A. A. Bakulin, *In Situ Observation Of Picosecond Polaron Self-Localisation In α-$Fe_2O_3$ Photoelectrochemical*

*Cells*, Nat. Commun. **10**, 3962 (2019).

[12] C. Boeglin, E. Beaurepaire, V. Halté, V. López-Flores, C. Stamm, N. Pontius, H. A. Dürr, and J. Y. Bigot, *Distinguishing The Ultrafast Dynamics Of Spin And Orbital Moments In Solids*, Nature **465**, 458 (2010).

[13] E. Morosan, H. W. Zandbergen, B. S. Dennis, J. W. G. Bos, Y. Onose, T. Klimczuk, A. P. Ramirez, N. P. Ong, and R. J. Cava, *Superconductivity In $Cu_XTiSe_2$*, Nat. Phys. **2**, 544 (2006).

[14] A. F. Kusmartseva, B. Sipos, H. Berger, L. Forró, and E. Tutiš, *Pressure Induced Superconductivity In Pristine 1T-TiSe₂*, Phys. Rev. Lett. **103**, 236401 (2009).

[15] K. C. Woo, F. C. Brown, W. L. McMillan, R. J. Miller, M. J. Schaffman, and M. P. Sears, *Superlattice Formation In Titanium Diselenide*, Phys. Rev. B **14**, 3242 (1976).

[16] J. A. Holy, K. C. Woo, M. V. Klein, and F. C. Brown, *Raman And Infrared Studies Of Superlattice Formation In TiSe₂*, Phys. Rev. B **16**, 3628 (1977).

[17] F. J. Di Salvo, D. E. Moncton, and J. V. Waszczak, *Electronic Properties And Superlattice Formation In The Semimetal TiSe₂*, Phys. Rev. B **14**, 4321 (1976).

[18] A. Kogar, M. S. Rak, S. Vig, A. A. Husain, F. Flicker, Y. Il Joe, L. Venema, G. J. MacDougall, T. C. Chiang, E. Fradkin, et al., *Signatures Of Exciton Condensation In A Transition Metal Dichalcogenide*, Science **358**, 1314 (2017).

[19] J.-P. Castellan, S. Rosenkranz, R. Osborn, Q. Li, K. E. Gray, X. Luo, U. Welp, G. Karapetrov, J. P. C. Ruff, and J. van Wezel, *Chiral Phase Transition In Charge Ordered 1T−TiSe2*, Phys. Rev. Lett. **110**, 196404 (2013).

[20] S.-Y. Xu, Q. Ma, Y. Gao, A. Kogar, G. Zong, A. M. M. Valdivia, T. H. Dinh, S.-M. Huang, B. Singh, C.-H. Hsu, et al., *Optical Detection And Manipulation Of Spontaneous Gyrotropic Electronic Order In A Transition-Metal Dichalcogenide Semimetal*, Nat. 2020 5787796 **578**, 545 (2019).

[21] D. Qian, D. Hsieh, L. Wray, E. Morosan, N. L. Wang, Y. Xia, R. J. Cava, and M. Z. Hasan, *Emergence Of Fermi Pockets In A New Excitonic Charge-Density-Wave Melted Superconductor*, Phys. Rev. Lett. **98**, 117007 (2007).

[22] M. D. Watson, O. J. Clark, F. Mazzola, I. Marković, V. Sunko, T. K. Kim, K. Rossnagel, and P. D. C. King, *Orbital- And Kz -Selective Hybridization Of Se 4p And Ti 3d States In The Charge Density Wave Phase Of TiSe2*, Phys. Rev. Lett. **122**, 076404 (2019).

[23] H. Cercellier, C. Monney, F. Clerc, C. Battaglia, L. Despont, M. G. Garnier, H. Beck, P. Aebi, L. Patthey, H. Berger, et al., *Evidence For An Excitonic Insulator Phase In 1T-TiSe$_2$*, Phys. Rev. Lett. **99**, 146403 (2007).

[24] T. Pillo, J. Hayoz, H. Berger, and F. Lévy, *Photoemission Of Bands Above The Fermi Level: The Excitonic Insulator Phase Transition*, Phys. Rev. B - Condens. Matter Mater. Phys. **61**, 16213 (2000).

[25] M. D. Watson, A. M. Beales, and P. D. C. King, *On The Origin Of The Anomalous Peak In The Resistivity Of TiSe$_2$*, Phys. Rev. B **99**, 195142 (2019).

[26] G. Li, W. Z. Hu, D. Qian, D. Hsieh, M. Z. Hasan, E. Morosan, R. J. Cava, and N. L. Wang, *Semimetal-To-Semimetal Charge Density Wave Transition In 1T-TiSe$_2$*, Phys. Rev. Lett. **99**, 027404 (2007).

[27] M. L. Mottas, T. Jaouen, B. Hildebrand, M. Rumo, F. Vanini, E. Razzoli, E. Giannini, C. Barreteau, D. R. Bowler, C. Monney, et al., *Semimetal-To-Semiconductor Transition And Charge-Density-Wave Suppression In 1T-TiSe$_{2-X}$S$_X$ Single Crystals*, Phys. Rev. B **99**, 155103 (2019).

[28] H. P. Hughes, *Structural Distortion In TiSe$_2$ And Related Materials-A Possible Jahn-Teller Effect?*, J. Phys. C Solid State Phys. **10**, L319 (1977).

[29] W. Y. Liang, in *Intercalation Layer. Mater.* (Springer, Boston, MA, 1986), pp. 31–73.

[30] C. Monney, C. Battaglia, H. Cercellier, P. Aebi, and H. Beck, *Exciton Condensation Driving The Periodic Lattice Distortion Of 1T-TiSe$_2$*, Phys. Rev. Lett. **106**, 106404 (2011).

[31] T. Rohwer, S. Hellmann, M. Wiesenmayer, C. Sohrt, A. Stange, B. Slomski, A. Carr, Y. Liu, L. M. Avila, M. Kalläsignne, et al., *Collapse Of Long-Range Charge Order Tracked By Time-Resolved Photoemission At High Momenta*, Nature **471**, 490 (2011).

[32] T. Jaouen, M. Rumo, B. Hildebrand, M.-L. Mottas, C. W. Nicholson, G. Kremer, B. Salzmann, F. Vanini, C. Barreteau, E. Giannini, et al., *Unveiling The Semimetallic Nature Of 1T-TiSe2 By Doping Its Charge Density Wave*, arXiv: 2006.08983 (2019).

[33] C. Monney, G. Monney, P. Aebi, and H. Beck, *Electron-Hole Instability In 1T-TiSe$_2$*, New J. Phys. **14**, 075026 (2012).

[34] M. Hellgren, J. Baima, R. Bianco, M. Calandra, F. Mauri, and L. Wirtz, *Critical Role Of The Exchange Interaction For The Electronic Structure And Charge-Density-Wave Formation In TiSe$_2$*, Phys. Rev. Lett. **119**, 176401 (2017).

[35] C. Monney, H. Cercellier, F. Clerc, C. Battaglia, E. F. Schwier, C. Didiot, M. G. Garnier, H. Beck, P. Aebi, H. Berger, et al., *Spontaneous Exciton Condensation In 1T-TiSe2: BCS-Like Approach*, Phys. Rev. B **79**, 045116 (2009).

[36] G. Monney, C. Monney, B. Hildebrand, P. Aebi, and H. Beck, *Impact Of Electron-Hole Correlations On The 1T-TiSe$_2$ Electronic Structure*, Phys. Rev. Lett. **114**, 086402 (2015).

[37] S. Mathias, S. Eich, J. Urbancic, S. Michael, A. V. Carr, S. Emmerich, A. Stange, T. Popmintchev, T. Rohwer, M. Wiesenmayer, et al., *Self-Amplified Photo-Induced Gap Quenching In A Correlated Electron Material*, Nat. Commun. **7**, 12902 (2016).

[38] T. Kaneko, Y. Ohta, and S. Yunoki, *Exciton-Phonon Cooperative Mechanism Of The Triple-Q Charge-Density-Wave And Antiferroelectric Electron Polarization In TiSe$_2$*, Phys. Rev. B **97**, 155131 (2018).

[39] J. van Wezel, P. Nahai-Williamson, and S. S. Saxena, *Exciton-Phonon-Driven Charge Density Wave In TiSe$_2$*, Phys. Rev. B **81**, 165109 (2010).

[40] J. Van Wezel, P. Nahai-Williamson, and S. S. Saxena, *An Alternative Interpretation Of Recent ARPES Measurements On TiSe$_2$*, EPL **89**, 47004 (2010).

[41] C. Chen, B. Singh, H. Lin, and V. M. Pereira, *Reproduction Of The Charge Density Wave Phase Diagram In 1T-TiSe$_2$ Exposes Its Excitonic Character*, Phys. Rev. Lett. **121**, 226602 (2018).

[42] V. Scagnoli, U. Staub, A. M. Mulders, M. Janousch, G. I. Meijer, G. Hammerl, J. M. Tonnerre, and N. Stojic, *Role Of Magnetic And Orbital Ordering At The Metal-Insulator Transition In NdNiO3*, Phys. Rev. B **73**, 100409 (2006).

[43] U. Staub, G. I. Meijer, F. Fauth, R. Allenspach, J. G. Bednorz, J. Karpinski, S. M. Kazakov, L. Paolasini, and F. D'Acapito, *Direct Observation Of Charge Order In An Epitaxial [Formula Presented] Film*, Phys. Rev. Lett. **88**, 4 (2002).

[44] G. Ghiringhelli, M. Le Tacon, M. Minola, S. Blanco-Canosa, C. Mazzoli, N. B. Brookes, G. M. De Luca, A. Frano, D. G. Hawthorn, F. He, et al., *Long-Range Incommensurate Charge Fluctuations In (Y,Nd)Ba$_2$Cu$_3$O$_{6+x}$*, Science **337**, 821 (2012).

[45] V. Scagnoli, U. Staub, Y. Bodenthin, R. A. De Souza, M. García-Fernández, M. Garganourakis, A. T. Boothroyd, D. Prabhakaran, and S. W. Lovesey, *Observation Of Orbital Currents In CuO*, Science **332**, 696 (2011).

[46] Y. Murakami, J. P. Hill, D. Gibbs, M. Blume, I. Koyama, M. Tanaka, H. Kawata, T. Arima, Y. Tokura, K. Hirota, et al., *Resonant X-Ray Scattering From Orbital Ordering In LaMnO$_3$*, Phys.

Rev. Lett. **81**, 582 (1998).

[47] T. Huber, S. O. Mariager, A. Ferrer, H. Schäfer, J. A. Johnson, S. Grübel, A. Lübcke, L. Huber, T. Kubacka, C. Dornes, et al., *Coherent Structural Dynamics Of A Prototypical Charge-Density-Wave-To-Metal Transition*, Phys. Rev. Lett. **113**, 026401 (2014).

[48] V. Esposito, L. Rettig, E. M. Bothschafter, Y. Deng, C. Dornes, L. Huber, T. Huber, G. Ingold, Y. Inubushi, T. Katayama, et al., *Dynamics Of The Photoinduced Insulator-To-Metal Transition In A Nickelate Film*, Struct. Dyn. **5**, 064501 (2018).

[49] S. de Jong, R. Kukreja, C. Trabant, N. Pontius, C. F. Chang, T. Kachel, M. Beye, F. Sorgenfrei, C. H. Back, B. Bräuer, et al., *Speed Limit Of The Insulator–Metal Transition In Magnetite*, Nat. Mater. **12**, 882 (2013).

[50] E. Möhr-Vorobeva, S. L. Johnson, P. Beaud, U. Staub, R. De Souza, C. Milne, G. Ingold, J. Demsar, H. Schaefer, and A. Titov, *Nonthermal Melting Of A Charge Density Wave In TiSe2*, Phys. Rev. Lett. **107**, 036403 (2011).

[51] M. Porer, M. Fechner, M. Kubli, M. J. Neugebauer, S. Parchenko, V. Esposito, A. Narayan, N. A. Spaldin, R. Huber, M. Radovic, et al., *Ultrafast Transient Increase Of Oxygen Octahedral Rotations In A Perovskite*, Phys. Rev. Res. **1**, 012005 (2019).

[52] P. Beaud, S. L. Johnson, E. Vorobeva, U. Staub, R. A. De Souza, C. J. Milne, Q. X. Jia, and G. Ingold, *Ultrafast Structural Phase Transition Driven By Photoinduced Melting Of Charge And Orbital Order*, Phys. Rev. Lett. **103**, 155702 (2009).

[53] T. Ishikawa, H. Aoyagi, T. Asaka, Y. Asano, N. Azumi, T. Bizen, H. Ego, K. Fukami, T. Fukui, Y. Furukawa, et al., *A Compact X-Ray Free-Electron Laser Emitting In The Sub-Ångström Region*, Nat. Photonics **6**, 540 (2012).

[54] T. Katayama, S. Owada, T. Togashi, K. Ogawa, P. Karvinen, I. Vartiainen, A. Eronen, C. David, T. Sato, K. Nakajima, et al., *A Beam Branching Method For Timing And Spectral Characterization Of Hard X-Ray Free-Electron Lasers*, Struct. Dyn. **3**, 034301 (2016).

[55] K. Nakajima, Y. Joti, T. Katayama, S. Owada, T. Togashi, T. Abe, T. Kameshima, K. Okada, T. Sugimoto, M. Yamaga, et al., *Software For The Data Analysis Of The Arrival-Timing Monitor At SACLA*, J. Synchrotron Radiat. **25**, 592 (2018).

[56] M. Porer, U. Leierseder, J.-M. Ménard, H. Dachraoui, L. Mouchliadis, I. E. Perakis, U. Heinzmann, J. Demsar, K. Rossnagel, and R. Huber, *Non-Thermal Separation Of Electronic And Structural Orders In A Persisting Charge Density Wave*, Nat. Mater. **13**, 857 (2014).

[57] H. Hedayat, C. J. Sayers, D. Bugini, C. Dallera, D. Wolverson, T. Batten, S. Karbassi, S. Friedemann, G. Cerullo, J. van Wezel, et al., *Excitonic And Lattice Contributions To The Charge Density Wave In 1T-TiSe$_2$ Revealed By A Phonon Bottleneck*, Phys. Rev. Res. **1**, 023029 (2019).

[58] H. Ueda, M. Porer, J. R. L. Mardegan, S. Parchenko, N. Grurung, F. Fabrizi, M. Ramakrishnan, L. Boie, M. J. Neugebauer, B. Burganov, et al., *Correlation Between Electronic And Structural Orders In 1T-TiSe2*, arXiv: 2006.08983 (2020).

[59] S. Kitou, A. Nakano, S. Kobayashi, K. Sugawara, N. Katayama, N. Maejima, A. Machida, T. Watanuki, K. Ichimura, S. Tanda, et al., *Effect Of Cu Intercalation And Pressure On Excitonic Interaction In 1T-TiSe$_2$*, Phys. Rev. B **99**, 104109 (2019).

[60] Note that the observed dynamics are approximately three times slower than the experimental time-resolution (~50 fs, FWHM - see Experimental section), ruling out immediate effects of hot-carrier injection [30] on the Se 4p orbital geometry and the hole population.

[61] P. Beaud, A. Caviezel, S. O. Mariager, L. Rettig, G. Ingold, C. Dornes, S. W. Huang, J. A. Johnson, M. Radovic, T. Huber, et al., *A Time-Dependent Order Parameter For Ultrafast Photoinduced Phase Transitions*, Nat. Mater. **13**, 923 (2014).

[62] M. Porer, L. Rettig, E. M. Bothschafter, V. Esposito, R. B. Versteeg, P. H. M. van Loosdrecht, M. Savoini, J. Rittmann, M. Kubli, G. Lantz, et al., *Correlations Between Electronic Order And Structural Distortions And Their Ultrafast Dynamics In The Single-Layer Manganite Pr$_{0.5}$Ca$_{1.5}$MnO$_4$*, Phys. Rev. B **101**, 075119 (2020).

[63] S. C. Bayliss and W. Y. Liang, *Reflectivity, Joint Density Of States And Band Structure Of Group IVb Transition-Metal Dichalcogenides*, J. Phys. C Solid State Phys. **18**, 3327 (1985).

[64] Note that the absorbed fluence of 33 μJ/cm$^2$ corresponds to excitation densities of only 2.4 x 10$^{-3}$ photons per Ti-atom, rendering the contribution of hole depopulation on the scattering intensity negligible.

[65] N. Wakabayashi, H. G. Smith, K. C. Woo, and F. C. Brown, *Phonons And Charge Density Waves In 1T-TiSe$_2$*, Solid State Commun. **28**, 923 (1978).

[66] M. Holt, P. Zschack, H. Hong, M. Y. Chou, and T. C. Chiang, *X-Ray Studies Of Phonon Softening In TiSe$_2$*, Phys. Rev. Lett. **86**, 3799 (2001).

[67] F. Weber, S. Rosenkranz, J. P. Castellan, R. Osborn, G. Karapetrov, R. Hott, R. Heid, K. P. Bohnen, and A. Alatas, *Electron-Phonon Coupling And The Soft Phonon Mode In TiSe$_2$*, Phys. Rev. Lett. **107**, 266401 (2011).

[68] R. Yusupov, T. Mertelj, V. V. Kabanov, S. Brazovskii, P. Kusar, J. H. Chu, I. R. Fisher, and D. Mihailovic, *Coherent Dynamics Of Macroscopic Electronic Order Through A Symmetry Breaking Transition*, Nat. Phys. **6**, 681 (2010).

[69] Also in this case, we cannot find agreement of model and all experimental traces when using the literature absorption coefficient.

[70] D. H. Torchinsky, F. Mahmood, A. T. Bollinger, I. Božović, and N. Gedik, *Fluctuating Charge-Density Waves In A Cuprate Superconductor*, Nat. Mater. **12**, 387 (2013).

[71] S. Hellmann, T. Rohwer, M. Kalläne, K. Hanff, C. Sohrt, A. Stange, A. Carr, M. M. Murnane, H. C. Kapteyn, L. Kipp, et al., *Time-Domain Classification Of Charge-Density-Wave Insulators*, Nat. Commun. **3**, 1069 (2012).

[72] S. Manzeli, D. Ovchinnikov, D. Pasquier, O. V. Yazyev, and A. Kis, *2D Transition Metal Dichalcogenides*, Nat. Rev. Mater. **2**, 17033 (2017).

[73] M. Hepting, R. J. Green, Z. Zhong, M. Bluschke, Y. E. Suyolcu, S. Macke, A. Frano, S. Catalano, M. Gibert, R. Sutarto, et al., *Complex Magnetic Order In Nickelate Slabs*, Nat. Phys. **14**, 1097 (2018).

[74] R. Ramesh and D. G. Schlom, *Creating Emergent Phenomena In Oxide Superlattices*, Nat. Rev. Mater. **4**, 257 (2019).

[75] G. Rainò, M. A. Becker, M. I. Bodnarchuk, R. F. Mahrt, M. V. Kovalenko, and T. Stöferle, *Superfluorescence From Lead Halide Perovskite Quantum Dot Superlattices*, Nature **563**, 671 (2018).

[76] Z. Luo, T. P. Dao, A. Hrabec, J. Vijayakumar, A. Kleibert, M. Baumgartner, E. Kirk, J. Cui, T. Savchenko, G. Krishnaswamy, et al., *Chirally Coupled Nanomagnets*, Science **363**, 1435 (2019).

[77] S. H. Skjærvø, C. H. Marrows, R. L. Stamps, and L. J. Heyderman, *Advances In Artificial Spin Ice*, Nat. Rev. Phys. **2**, 13 (2020).

# Popular Summary

Unconventional material properties, such as superconductivity or charge-density-wave formation, arise through the simultaneous condensation of electronic and structural order. The phase transitions into and out of such unconventional states are commonly conceived to be kinetically dominated by electronic phenomena. Here, we show the opposite: suppression of the structural order can outpace macroscopic electronic interactions and hence govern melting of the charge-density-wave phase in 1T-TiSe$_2$.

Our findings are based on experimental data from ultrafast pump-probe non-resonant and resonant X-ray diffraction measurements, performed at a free electron laser facility. In this experiment, we are able to track both the structural and the (electronic) orbital order as a function of time after laser excitation and with increasing laser strength. We observe a distinct energy regime in which complete suppression of both orders occurs more than one order of magnitude below the suggested excitonic binding energy and drastically faster than expected from a purely electronic charge-screening process. In this regime, our findings strongly suggest a structurally assisted breakup of excitonic correlations, as structural destabilization outpaces the macroscopic screening of excitonic interactions.

The here made observation that melting of such unconventional states can occur via different phase transition pathways may have a significant impact on all material systems with multiple condensed orders. This is particularly drastic for systems with strong electron-phonon coupling such as cuprates, other dichalcogenides, (complex) oxides as well as superlattice and spin-ice systems that are highly debated as promising candidates for superconducting or magneto-optical applications.